\documentclass{ifacconf}

\usepackage{graphicx}      
\usepackage{natbib}        
\usepackage[english]{babel}
\usepackage{amsmath}
\usepackage{epstopdf}
\usepackage{flushend}
\usepackage{algorithm}
\usepackage{algpseudocode}
\usepackage{caption}

\begin{document}
\begin{frontmatter}

\title{Location Information Sharing Using Software Defined Radio in Multi-UAV Systems} 

\author[First]{Mehmet Kaan EROL} 
\author[Second]{Eyüp Emre ÜLKÜ}

\address[First]{Department of Computer Engineering, Faculty of Technology, Marmara University (e-mail: kaanerol@ marun.edu.tr).}
\address[Second]{Department of Computer Engineering, Faculty of Technology, Marmara University (e-mail: emre.ulku@marmara.edu.tr)}

\begin{abstract}                
SDR (Software Defined Radio) provides flexible, reproducible, and longer-lasting radio tools for military and civilian wireless communications infrastructure. SDR is a radio communication system whose components are implemented as software. This study aims to establish multi-channel wireless communication with FANET between two SDRs to share location information and examine it in a realistic test environment. We used multi-channel token circulation as a channel access protocol and GNU Radio platform for SDR software development. The structures of the communication layer, including the protocols, communication systems, and network structures suggested in the studies in the literature, are generally tested in the simulation environment. The simulation environment provides researchers with fast and easy development and testing, but disadvantages exist. These cause a product to be isolated from hardware, software, and cost effects encountered while developing and environmental factors affecting the communication channel while testing. Another contribution of the study is to present the developed block diagrams and codes as clear and reproducible. The developed software and block diagrams are available at github.com/knrl/uav-in-802.11-gnuradio.
\end{abstract}


\end{frontmatter}

\section{Introduction}
Unmanned Aerial Vehicles (UAVs) are a technology that has found applications in both military and various civilian sectors, owing to features such as ease of use, low cost, and high mobility \cite{shakhatreh2019unmanned}. These vehicles can be adapted for different purposes across various sectors by altering their designs or payloads. Their versatility in mobility and payload diversity further enhances the functionality of these vehicles. Configurations involving multiple UAVs, which consist of more than one UAV, are capable of performing more complex and challenging missions more effectively \cite{gupta2015survey}. Additionally, they possess a small radar cross-section and are easily scalable for a variety of missions \cite{bekmezci2013flying}. Zhou et al. presented a more manageable model for multi-UAV applications by dividing them into five layers: the decision-making layer, route planning layer, control layer, communication layer, and application layer. This study focuses on the communication layer of multi-UAV systems \cite{zhou2020uav}.

While multi-UAV systems can efficiently perform challenging and complex missions, ensuring that these systems act in coordination and cooperation is a highly challenging process. Particularly, it is crucial to ensure seamless communication between UAVs. This study presents a multi-token circulation-based method to facilitate communication in multi-UAV systems. Utilizing the Flying Ad-Hoc Network (FANET) structure, the method enables fast and reliable communication among UAVs. The study established a realistic multi-UAV communication system using the Orthogonal Frequency Division Multiplexing (OFDM) technique for communication, Software Defined Radio (SDR) for the test environment, and the GNU Radio platform for software development. While simulation environments provide researchers with fast and convenient means for development and testing of communication layer-related applications, they also have various disadvantages. One such disadvantage is their isolation from the hardware, software, and cost factors involved in product development, as well as from the environmental factors that affect the communication channel during testing.

The most fundamental challenge in designing multi-UAV systems is communication \cite{bekmezci2013flying}. There are numerous issues that need addressing to enable multi-UAVs to operate within reliable stable, context-specific network structures \cite{gupta2015survey}. Effective communication is crucial for these systems to successfully and efficiently complete their missions. Whether it's for the swarm to move in harmony or for each individual vehicle to perform its designated task, seamless communication and information transfer between the UAVs are essential.

Successful mission fulfillment in multi-UAV systems hinges on the effective transfer and accurate acquisition of information \cite{zhou2020uav}. As the number of UAVs in a multi-UAV system increases, the complexity of communication challenges escalates. For a system where each UAV operates at varying speeds in three dimensions, powered only by its battery capacity, establishing an efficient network structure and communication is crucial. One approach to ensure system-wide communication is to connect all UAVs to a ground station via satellite or other infrastructure \cite{bekmezci2015flying}. However, reliance on infrastructure or satellite-based communication architectures can limit the operational capacity of multi-UAV systems \cite{bekmezci2015flying}. Frew et al. highlighted the operational network challenges that multi-UAV systems will face in future applications \cite{frew2009networking}, summarizing these challenges into three main areas: data transmission, network connectivity, and service discovery \cite{frew2009networking}. Due to their effectiveness in addressing these three key issues, multi-UAV systems often prefer ad-hoc network solutions over others \cite{george2011search}.

\subsection{FANET}
Ad-hoc networks are a distributed wireless network structure that allows communication between nodes without the need for infrastructure \cite{sun2011bordersense}.  Ad-hoc networks exhibit various specialized structures, including Mobile Ad-Hoc Networks (MANET), Vehicular Ad-Hoc Networks (VANET), and Flying Ad-Hoc Networks (FANET). Compared to other structures, FANET boasts high node mobility, significant node-to-node distances, and peer-to-peer communication, facilitating coordination and cooperation \cite{gupta2015survey,bekmezci2013flying}. With advancements in UAV technology and decreasing costs, the utilization of multi-UAV systems is becoming increasingly prevalent. As the usage of multi-UAV systems continues to rise, the application areas of the FANET structure are expanding." The widespread use of the FANET structure has brought new challenges such as multipath propagation, shadowing, traffic load balancing, congestion, mobility, and high error rate \cite{chughtai2021ieee}. In the literature, studies also focus on these problems. Khan et al. introduced communication architectures suitable for FANET discussed various routing protocols for FANET, and showed the open research topics of existing protocols \cite{khan2017flying}. The communication architecture represents the structure that defines how the information is transmitted between the UAV and the ground station. The architecture used for FANET should be decentralized, robust, and have real-time performance against variable distances, and it is extremely challenging to establish a communication architecture with these features \cite{gupta2015survey, khan2017flying}. Routing protocols, as well as the architecture of the network, play an important role in improving the performance of ad hoc networks \cite{singh2015experimental}. Singh et al. analyzed AODV, DSDV, and OLSR routing protocols for FANET networks according to packet delivery rate, end-to-end delay, and throughput parameters \cite{singh2015experimental}. In multi-uav communication research, testbeds are also crucial to contribute. Bekmezci et al. proposed a test environment that supports the FANET structure, suitable for price performance, and can be easily re-applied, and the test environment was applied on UAVs with Raspberry Pi 3 and wifi modules \cite{bekmezci2015flying}. In various studies, applications including the decision process, have been made in the FANET-based Multi-UAV system \cite{chaumette2011carus, ben2008distributed, alshbatat2012cooperative}. Various attacks against FANETs were analyzed by Ceviz et al., 3D network scenarios were constructed in the simulator environment, and an artificial neural network application was investigated from the obtained data set \cite{tacs2021nesnelerin}. Most of these studies are tested and evaluated with simulators for ease of application. Simulators cannot adequately model physical effects, and a test environment can be created to create an application environment closest to the real world \cite{tabrizi2019ad}. For this purpose, Software Defined Radio (SDR) is preferred in the literature.

\subsection{Software Defined Radio}
SDRs are preferred in test environment setup and tactical uses due to their ease of use and cost-performance effectiveness. SDR provides flexible, retrofit, and longer-lasting radio tools for military and civilian wireless communications infrastructure \cite{ulversoy2010software}. Progress in the field of SDR has led to an increase in protocol development and application areas with greater emphasis on programmability, flexibility, portability, and energy efficiency in cellular, WiFi, and M2M communication \cite{akeela2018software}. SDR allows for reduced development costs for end users and simplifies the implementation of communication protocols. An SDR can seamlessly communicate with or act as a bridge between multiple, incompatible radios. Unlike specialized internal circuit designs, SDR devices can operate in wide frequency ranges and bandwidths and can be developed as software. SDRs can be used as an RF front-end in a study where multi-UAV system application is desired. RF front ends convert the I and Q samples in binary form to radio frequency at the desired center frequency, bandwidth, and power. We aimed to use these radios to establish the communication system, which is an ad-hoc network. Tabrizi et al. developed an ad hoc indoor test environment using SDR and evaluated the test environment according to carrier frequency, transmission rate, carrier threshold size, packet size, transmission, and reception power parameters \cite{tabrizi2019ad}. Shi et al. used SDR in two different applications, mobility, and beamforming, and shared software and hardware test environments \cite{shi2018building}.

\subsection{Location Information Sharing}
This study aims to implement token circulation to share location information among UAVs with multi-channel communication. Multiple UAV systems have dynamic topologies. This leads to various problems such as overhearing, coordination, and collision as the number of nodes increases. In the literature, it has been proposed to share location information between UAVs using a FANET structure and to distribute multi-token keys for collision avoidance \cite{ulku2016multi, bekmezci2015location,ulku2019sharing}. Unlike previous studies, we created a realistic test environment with a dynamic topology in simulation and demonstrated the impact of varying bit error rate (BER) values. Ulku et al. proposed a multi-token circulation protocol over a common channel in their study \cite{ulku2016multi}. In another study, they proposed a multi-channel protocol assuming that communication is done over more than one channel and that appropriate hardware is provided to prevent conflicts between tokens \cite{ulku2019sharing}. 

This study aims to establish a completely realistic test environment by eliminating assumptions, distributing tokens through multiple channels, sharing location information, and analyzing environmental impacts. Another contribution of the study is that the block diagrams and codes developed to contribute to future studies in this field are presented in an open and easily reproducible manner. Based on the literature review conducted, no studies were identified that depict a technique for distributing multiple tokens on a single channel or circulating multiple tokens across multiple channels.  Some publications propose a multi-token circulation protocol and produce common or multi-channel models for multi-token circulation. However, these publications were evaluated in a simulation environment with various assumptions. Although SDRs are widely used in the literature in UAV applications, no study was found in which the collision avoidance protocol was applied. In addition, although FANET is widely preferred in multi-UAV studies, very few studies create a test environment with SDR. In this study, a realistic test environment was established. Another contribution of the study is sharing all the developed block diagrams and codes as reproducible.

\begin{table*}[h]
\centering
\caption{IEEE 802.11p Coding Structures and Values \cite{lin2012comparison}}
\label{table:802_11p}
\begin{tabular}{|c|c|c|c|c|c|}
\hline
MCS & Modulation, Coding Rate & Total Rate (Mb/s) & Min. SINR [dB] & Range [m] (PL) & Duration [s] \\ \hline
1   & BPSK, 1/2               & 3.0               & 10.0           & 223            & 848          \\ \hline
2   & BPSK, 3/4               & 4.5               & 11.0           & 210            & 584          \\ \hline
3   & QPSK, 1/2               & 6.0               & 13.0           & 188            & 448          \\ \hline
4   & QPSK, 3/4               & 9.0               & 15.0           & 167            & 312          \\ \hline
5   & 16QAM, 1/2              & 12.0              & 18.0           & 141            & 248          \\ \hline
6   & 16QAM, 3/4              & 18.0              & 22.0           & 112            & 176          \\ \hline
7   & 64QAM, 1/2              & 24.0              & 26.0           & 89             & 144          \\ \hline
8   & 64QAM, 3/4              & 27.0              & 27.0           & 84             & 136          \\ \hline
\end{tabular}
\end{table*}

\section{Methods}
This section mentions the technologies used in this study, the technical details of the study, and the application process. Information was given about some of the difficulties encountered during the implementation process and how they were resolved. The structure of SDR, the RF front-end structure, USRPs, and the GNU Radio platform used in the software development process were mentioned in detail.

\subsection{Software Defined Radio}
SDRs are radios that can be performed by software. SDR facilitates adaptable and versatile radio system creation, where most signal processing tasks occur through software rather than relying on dedicated hardware elements. It allows to perform different functions on the same platform at different times and to implement various baseband radio features (modulation, error verification coding, etc.) in software \cite{wyglinski2013digital}. Due to these features, SDR systems are preferred for research and development purposes.

\subsection{The Universal Software Radio Peripheral (USRP)}
USRP is the well-known open-source SDR hardware platform developed by Ettus Research LLC \cite{wyglinski2013digital}. Three generations and dozens of types of USRP hardware have been released so far \cite{ettus2015universal}. In this study, the 3rd generation USRP Embedded Series 310 model was used. USRP is fully supported by the USRP Hardware Driver™ (UHD) API. UHD is an open-source driver software under the GPLv3 license that allows porting designs and applications to USRPs. It is a library that runs on a general-purpose processor and communicates with and controls the entire USRP device family \cite{webref4}. It provides the necessary control to carry RF wave samples to or from the USRP hardware and controls various radio parameters (sample rate, center frequency, amplifications) \cite{webref4}.

\subsection{GNU Radio}
GNU Radio is an open-source software development kit and framework licensed under the General Public License (GPL). GNU Radio includes software-defined radio, signal processing, and communications packages. Development can be done through the GNU Radio interface or its framework. The system can be built with hierarchical blocks and external blocks. Each block performs a single operation, which can be blocked for signal processing (filtering, channel coding, synchronization, editing), communication, hardware configuration, or communication. Connections between blocks also represent the data flow, governing how it is transferred. These blocks were developed in C++ and linked to each other at a high level with Python. When an external block is created, it is developed with C++ or a hierarchical block can be created by connecting existing blocks with Python. Thus, the developer can create real-time, high-efficiency radio systems in an easy-to-use, rapid development environment \cite{webref5}.

With GNU Radio, an application can be evaluated in 3 stages: establishing connections between Python and C++ signal blocks, the signal blocks themselves, and controlling the signal flow with the scheduler. The scheduler uses Python's internal threads to control "start," "stop," or "wait" operations \cite{webref6}. Architecturally, GNU Radio has a backpressure-driven parallel signal processing architecture \cite{webref7}.

\subsection{Communication System}
For effective ad-hoc UAV communication, UAVs must have compatible communication systems and protocols. We focused on developing a communication system compatible with the IEEE 802.11p standard using OFDM modulation. IEEE 802.11p is a wireless communication standard designed explicitly for use in vehicle environments, including UAVs \cite{bloessl2013towards}. It is based on the same technology as the more widely known IEEE 802.11a wireless standard. The important difference between 802.11p and 802.11a is the bandwidth, 10MHz, which helps the communication be more robust against fading and more resistant to multipath \cite{lin2012comparison}. Various studies have shown that the losses 802.11p are also significantly lower than those of 802.11a in both LOS and NLOS environments \cite{lin2012comparison}. In this study, a channel access protocol, which is more suitable for multiple UAV systems and allows efficient communication between multiple UAVs, was applied. In addition, the effects on the overall performance of the communication system were observed using various modulation techniques for OFDM's subcarriers.

\subsubsection{Modulation}
OFDM is a method used in digital communication to encode data into multiple subcarriers. During this coding, different modulation techniques can be used. This study used BPSK, QPSK, QAM-16, and QAM-64 techniques. BPSK is short for Binary Phase Shift Keying. It is a type of modulation used to transmit binary data over a communication channel in digital communication systems. In BPSK, the carrier signal is modulated by changing its phase 180 degrees for each binary symbol. QPSK, Quadrature Phase Shift Keying, is another phase modulation technique that transmits two bits per symbol. Instead of representing a QPSK symbol 0 or 1, it represents 00, 01, 10, or 11. QAM, Quadrature Amplitude Modulation, is an analog modulation method. It modulates two carrier signals using the amplitude shift keying (ASK) digital or the amplitude modulation (AM) analog modulation scheme. It transmits 2 bits per 1 symbol, while QAM-16 and QAM-64 use 4 and 8 bits per symbol, respectively.

\subsubsection{Throughput}
In communication, throughput refers to the amount of data that can be transmitted over a communication channel in a specified period of time. It is measured in bits per second (bps). Throughput is influenced by several factors, including the bandwidth of the channel, the quality of the channel, and the volume of data being transmitted. Additionally, the protocols used for data transmission also impact throughput. For instance, in a wired network, throughput may be constrained by the physical properties of the cable, whereas in a wireless network, it can be affected by factors such as distance, interference, and the number of devices sharing the same frequency band. In practice, the actual throughput experienced by users is often lower than the theoretical maximum, due to various factors like network congestion, packet loss, and latency.

Bandwidth refers to the maximum capacity of a communication channel to transmit data over a given period of time. It is typically measured in hertz (Hz) and is determined by the physical characteristics of the channel, such as its frequency range and noise levels. The data rate, also known as bit rate, represents the number of bits transmitted per unit of time over a communication channel. Commonly measured in bits per second (bps), it depends on factors like the modulation scheme used, the number of bits per symbol, and other transmission parameters. Throughput, in contrast, is the actual volume of data successfully transmitted over a communication channel within a certain period. It accounts for losses or inefficiencies in the transmission process, such as packet loss, retransmissions, or congestion. The maximum achievable throughput is constrained by factors like channel bandwidth and other elements in the transmission process, including error correction and congestion control. Table \ref{table:802_11p} examines the throughput of 802.11p under different modulation and coding schemes as investigated in \cite{lin2012comparison}, presenting the results in megabits per second (Mb/s).

\subsubsection{Channel Access Protocol}
Wireless channel access protocols manage communication between wireless devices in a shared radio frequency environment. These protocols determine how wireless devices access the available bandwidth and efficiently share the wireless channel. Without these protocols, wireless networks may experience interference, collisions, and poor network performance due to multiple devices on the same channel. Collision Avoidance with Carrier Sense Multiple Access (CSMA/CA): It is a widely used protocol in the IEEE 802.11 Wi-Fi standard \cite{bianchi1996performance}. This protocol requires wireless devices to listen to the channel and control other transmitting devices before transmitting their data. If a device detects other activity on the channel, it waits for a random amount before trying again. This random waiting helps avoid conflicts and ensures the channel is available to the sending device \cite{sabharwal2007opportunistic}.

\begin{figure}
    \centering
    \includegraphics[width=0.36\textheight]{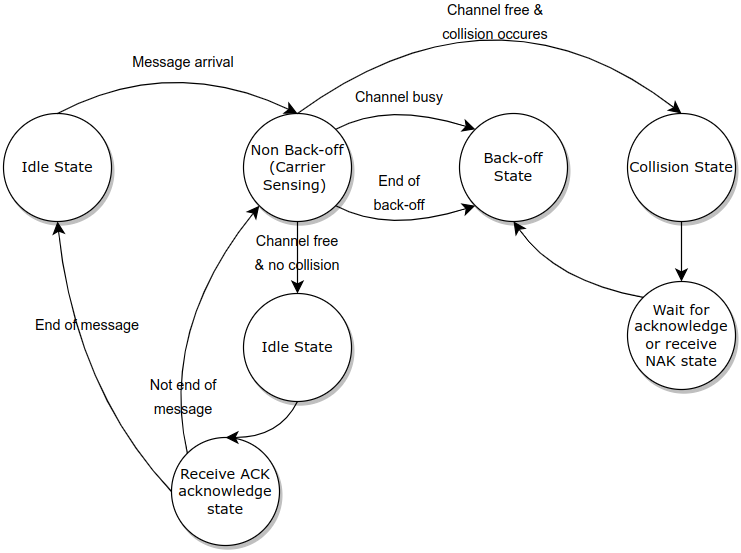}
    \caption{State transition diagram of CSMA \cite{state_diagram_article}}
    \label{fig:csma_state_diagram}
\end{figure}

The collision avoidance mechanism of CSMA/CA is based on virtual carrier detection (sense) whose state diagram is depicted in Figure \ref{fig:csma_state_diagram}. CSMA/CA uses a virtual carrier detection mechanism to avoid conflicts based on acknowledgment (ACK) packets sent by the receiving device. When a transmitting device sends data, it waits for an ACK packet from the receiving device. If the transmitting device does not receive an ACK packet within a certain time, it assumes a collision and retransmits the data after waiting for a random backoff period. CSMA/CA also uses a Request To Send (RTS) and Clear To Send (CTS) mechanism to improve collision avoidance further. When a device wants to transmit data, it sends an RTS to the receiving device. The receiving device responds with a CTS notifying the transmitting device that the channel is available for use. CSMA/CA is a wireless channel access protocol that uses a virtual carrier detection mechanism and a rollback algorithm to avoid collisions between wireless devices. The RTS/CTS mechanism is also used to improve collision avoidance further and provide efficient data transmission.

Although CSMA/CA is widely used today, more is needed in network structures designed for multiple UAV systems such as FANET. This is because in the multi-UAV system, collisions may increase, and efficiency may decrease with the increase of the number of devices. Additionally, in a multi-UAV system, there may be issues with hidden node problems where two UAVs may be out of range of each other but within range of a common ground station or access point. In such a case, CSMA/CA alone cannot prevent collisions because both UAVs may not be able to detect each other's broadcasts. For this reason, it is aimed to use the token circulation method proposed in \cite{ulku2016multi, bekmezci2015location,ulku2019sharing}.

\begin{figure*}[ht]
    \centering
    \includegraphics[width=0.85\textwidth]{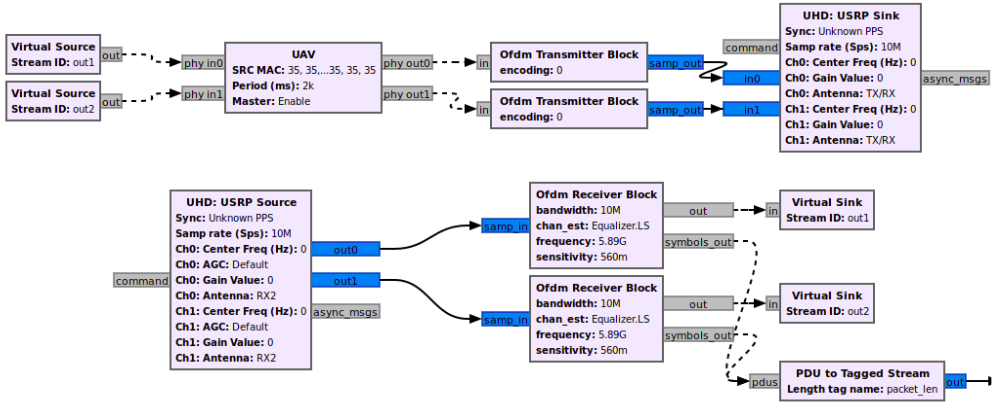}
    \caption{GNU Radio diagram of the system}
    \label{fig:gnu_diag}
\end{figure*}

\subsubsection{Multi-Token Circulation Protocol}
Multiple UAVs in FANET need each other's coordinated information to make smooth and coordinated flights. This section explains the multi-token circulation model developed to circulate location information between UAVs in a FANET. A token circulation-based approach can be used to solve this problem, but as the number of UAVs in the network increases, the geolocation packet grows, and more UAVs need to distribute the tokens. If there are fewer UAVs, more than a single token is needed. More tokens can be added to improve accuracy, but this can cause loss or latency issues due to token collisions. To avoid these problems, a multi-token circulation has been proposed where tokens circulate on a common channel and are routed through a second channel to avoid collisions \cite{ulku2019sharing}. 

This study applied the multi-token circulation protocol proposed in \cite{ulku2019sharing}. Algorithm \ref{alg:proposed_alg} depict how tokens are exchanged among UAVs in a FANET system. This algorithm shows that when a UAV receives a token, it updates the token with its own coordinate information. It also compares the token's data with its own stored data to ensure both contain the most recent information, using a counter variable to assist in this process. The counter increases as location information is updated, indicating the most current position data. Additionally, the concept of overhearing is integrated into this system. This means that UAVs not intended as the token's final recipient can still intercept and store the token's information if it's newer than what they have. However, they cannot alter the information without possessing the token. Overhearing is facilitated by the use of undirected antennas in wireless communication, which spread signals in all directions, allowing nearby UAVs to receive the information even if they are not the direct target. 

Another advantage of the proposed system is the usage of a "take-forward" approach. This approach is different from the "take-hold-forward" used in the classical ring token protocol. In the “receive-forward” approach, the incoming token is not kept during data transmission; it is sent directly, thus reducing the delay during transmission \cite{li2013token}. The token; consists of a token number, source address, destination address, and a table containing the location information of each UAV.

To avoid collisions in the multi-token circulation system, a second channel distributes small control packets such as Clear to Send (CTS), Request to Send (RTS) packets. Token holder UAVs use the adjacency matrix to determine which neighbors to send tokens to. Before sending the token, neighbors sending CTS and RTS packets are eliminated to avoid collisions. If only one neighbor remains after elimination, the token is sent to that neighbor. If there is no neighbor, the token is held until it is sent. If more than one neighbor is left, the source UAV is eliminated to ensure that the tokens are routed in different directions and that the data across the entire network is up to date.

\algdef{SE}[DOWHILE]{Do}{doWhile}{\algorithmicdo}[1]{\algorithmicwhile\ #1}%
\begin{algorithm}
\caption{Pseudocode of the proposed multi-token circulation method \cite{ulku2019sharing}}
\label{alg:proposed_alg}
\begin{algorithmic}
    \State $i \gets 1$
\While{UAVMove $\neq$ 0}
    \If{$i \leq \text{numberOfTokens}$}
        \State \textbf{label1:}
        \State $T \gets \text{token}(i)$
        \State $R \gets \text{randBerValue()}$
        \If{$T.\text{destination} == \text{thisUAV}$}
            \If{$R \leq \text{BER}$}
                \State $T \gets \text{updateToken}(\text{UAVCache}, T)$
                \State $\text{UAVCache} \gets \text{update}(\text{UAVCache}, T)$
            \Do
                \State $\text{updateAdjMatrix()}$
                \State $N \gets \text{findNeighbours}(\text{adjMatrix})$
                \State $S \gets \text{selectNextUAV}(N, \text{md})$
                \State $\text{sendToken}(T, S)$
                \State $R \gets \text{randBerValue()}$
            \doWhile{$R > \text{BER}$}
            \EndIf
        \EndIf
    \Else
        \State $\text{UAVCache} \gets \text{updateCache}(\text{UAVCache}, T)$
    \EndIf
\EndWhile
\end{algorithmic}
\end{algorithm}

\subsection{Proposed System}
The system designed in GNU Radio can be tested both in the simulation environment and with SDR hardware. Connecting the channel inputs and outputs to the USRP Sink and Source blocks will be sufficient. The system consists of three parts: transmitter, receiver, and UAV. UAV is a block in which a multi-token circulation is applied, and a UAV is simulated. In Figure \ref{fig:system_over} shows how the UAV block and the system work, the blocks are available at github.com/knrl/uav-in-802.11-gnuradio. The location information of the UAV is changed randomly over a certain period, ensuring its continuous movement. At the same time, it constantly listens to the channel and waits for the token; when the token arrives, the target is decided through the protocol described in 2.3.1, and the token is distributed. The UAV block has been created in such a way that it can be easily added and removed from the system. If a new UAV is wanted to be added to the system, connecting the block directly to the same channel will be sufficient. Each UAV block represents a UAV that moves randomly in a certain range, and the protocol implementation works the same for each.

In implementing the protocol, the control packets must be completed due to GNU Radio's inability to create network simulations. Control packages exist in small sizes and in short time intervals. In off-the-shelf network cards, these operations are performed at the kernel level and with high priority in operating systems. Since GNU Radio is not a network simulator, the delays are high, and the control packets cannot provide the desired functionality. Again, a similar problem has been observed in various studies \cite{bloessl2013ieee,mandke2007early,korakis2009cooperative}. Bloessl et al. In their studies, IEEE 802.11a/g/p was implemented; for similar reasons, the CSMA\textbackslash CA mechanism was not applied \cite{bloessl2013ieee}. Mandke et al. tried to prevent the timeout by increasing the packet size and delay \cite{mandke2007early}. However, such an application was avoided because it did not detract from the main focus of the study and the success of the protocol used has been demonstrated in their studies \cite{ulku2016multi, bekmezci2015location, ulku2019sharing}. This study aims to establish an SDR-based environment for token circulation and a multiple UAV system.

In Figure \ref{fig:gnu_diag}, the communication system for a single UAV is shown. In this system \cite{bloessl2013ieee}, OFDM modulation transmitter and receiver blocks developed by the IEEE 802.11p standard were used. USRP Sink and Source blocks are internal blocks that allow us to use hardware effectively through the UHD driver. The system's dashed lines represent asynchrdonous and continuous lines represent synchronous flow. Asynchronous communication between blocks is done with the Message Passing method provided by GNU Radio. The system has a loop; the UAVs package the location information mentioned in 2.3.1 and leave it as a token to the channel. Other UAVs listening to the same channel receive this packet and send the new token to the channel after passing it through the decision mechanism.

\section{Evaluation and Experimental Results}
The multi-UAV communication system described in Chapter 2 has been tested and evaluated in this chapter. The system details were tested on the GNU Radio platform, as mentioned in 2.4. The results were obtained by changing the parameters determined through the interface of GNU Radio one by one. The parameters used were determined as PDU size, number of UAV (node), modulation type, BER (Bit Error Rate), and SNR (Signal to Noise Ratio). The effects of these parameters on the received packet rate and yield were evaluated.

\subsection{Evaluation of System}
UAVs can perform different tasks or functions in a multi-UAV system, such as reconnaissance, surveillance, or shooting. In addition, UAVs can operate in dynamic and complex environments where communication links may be difficult due to interference, range limitations, and jamming, such as in urban areas, disaster areas, or military scenarios. An effective communication system should help overcome these challenges. For this purpose, the effect of noise from the relationship between SNR and BER, the effect of SNR on the received packet rate, and the effects of the number of UAVs on the transmitter efficiency were tested.

\subsection{Effect of Noise (Relationship between Throughput and SNR)}
\begin{figure}
    \centering
    \includegraphics[width=0.36\textheight]{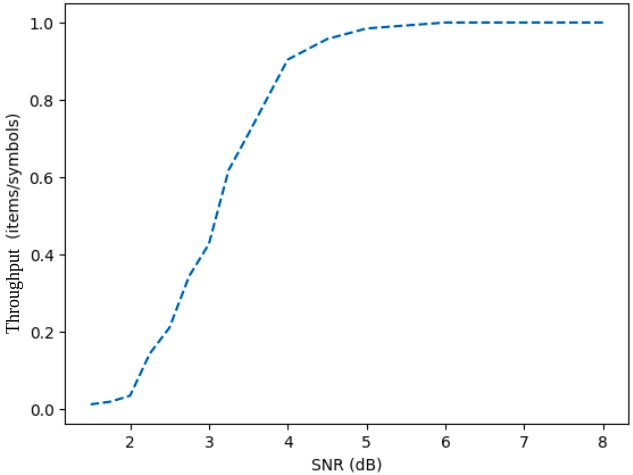}
    \caption{SNR and throughput relationship}
    \label{fig:snr_thr_rel}
\end{figure}

In a noisy channel, a higher SNR indicates that the desired signal is stronger than the background noise. This increased signal strength allows more accurate detection and decoding of transmitted data, resulting in a lower BER and higher throughput. The increase in throughput with higher SNR is due to the system being able to use more complex modulation schemes or error correction coding techniques as the SNR improves. The higher the SNR, the less important the efficiency improvement becomes. The returns decrease as the system approaches the theoretical capacity limit or other constraints imposed by the communication channel or system design. In addition, achievable throughput also depends on various factors such as modulation schemes, coding techniques, channel bandwidth, system design, and potential interference. Figure \ref{fig:snr_thr_rel} shows the efficiency and SNR relationship graph for the BPSK ½ coding of the proposed system. In this test, more than 100 samples were taken for each SNR value, and the average was plotted.

\subsection{Effect of SNR on Received Packet Rate}
\begin{figure}
    \centering
    \includegraphics[width=0.36\textheight]{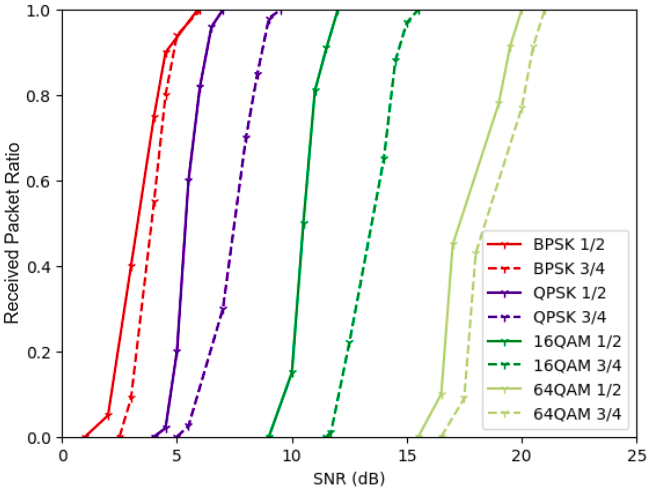}
    \caption{Received packet ratio}
    \label{fig:rcv_pck_rat}
\end{figure}

The higher the SNR, the better the forwarded packet rate. Sometimes the packet delivery rate measures the percentage of packets successfully received compared to the total number of packets transmitted. It indicates the reliability or effectiveness of the communication system in transmitting data without error or loss. In a noisy communication channel, a higher SNR means a stronger and more distinguishable signal than background noise. This increased signal strength allows better reception and decoding of transmitted packets, resulting in a higher transmission rate. A higher SNR provides a more significant margin for accurate detection and decoding of received packets, reducing the probability of error and increasing the probability of successful packet delivery. Channel conditions, modulation schemes, coding techniques, and system design can also affect the transmitted packet rate. In addition, the specific relationship between the SNR and the transmitted packet rate may vary depending on the communication system, protocol, and nature of the channel. This test was performed to characterize the system's performance under different SNR conditions and determine the reachable transmitted packet rate. Since the effect of SNR value varies according to different coding techniques, the system was tested with four different modulation techniques and two different coding for each. Figure \ref{fig:rcv_pck_rat} shows the relationship of the proposed system with the SNR for different encodings.

\subsection{Effects of Number of UAVs on Efficiency}
\begin{figure}
    \centering
    \includegraphics[width=0.36\textheight]{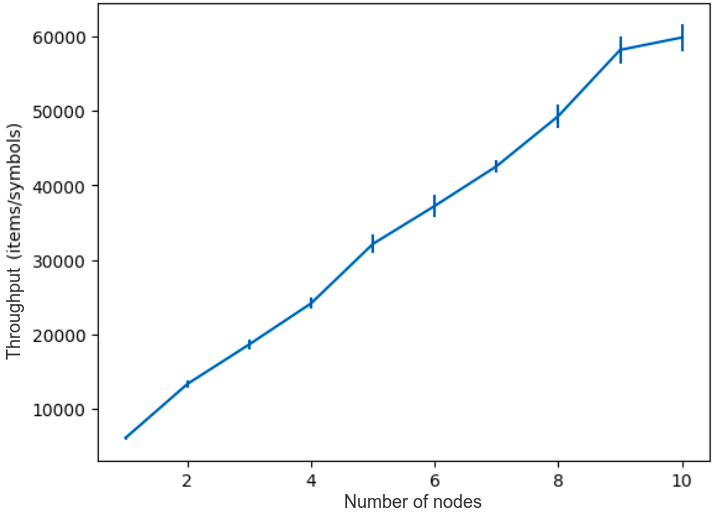}
    \caption{Number of node and throughput relationship}
    \label{fig:node_th_rel}
\end{figure}

Network performance can be affected when multiple nodes attempt to transmit data simultaneously or struggle for shared resources such as bandwidth or access to a medium. The relationship between the number of nodes and throughput can vary depending on the network architecture and the communication protocols used. In some cases, throughput may decrease due to increased contention and congestion as the number of nodes increases. This issue is generally in networks with limited capacity or shared resources, such as Ethernet or wireless networks. However, with proper network design, efficient routing algorithms, and congestion control mechanisms, it is possible to reduce the adverse effects of increasing the number of nodes and maintain or even improve throughput. This test analyzes network performance based on the number of nodes and desired throughput requirements. Here, throughput refers to the amount of data that can be successfully transmitted over a communication channel in a given time. The system was run 30 times for a few minutes for each node number, and its mean and standard deviations were taken. Figure \ref{fig:node_th_rel} shows the graph of this relationship. There is a direct relationship between the increase in the number of nodes and the number of successfully transmitted packets. However, this ratio starts to decrease after a while as the number of nodes increases due to problems such as congestion and overlap.

\section{Conclusion and Future Work}
In this study, the proposed multi-token circulation for the Multi-UAV system was implemented, and the necessary test environment was developed with software-defined radio. Similar studies in the literature, necessary platforms, and theoretical information were mentioned. As the communication technique in the system, four different coding scales were applied to the modulation of OFDM and its subcarriers. While developing the software environment, the GNU Radio platform, various blocks, and modules it offers were used. The work was built on gr-ieee802.11, an OOT module, as it satisfies the desired requirements \cite{bloessl2013towards}. This module has the necessary blocks for communication in 802.11a/b/p standards. Changes have been made to some blocks of this module, and a "UAV System" block has been added. After applying the desired blocks, the proposed multi-token circulation algorithm on the 802.11p wireless communication standard was applied. The proposed algorithm did not test in a network simulation since previous studies successfully tested it \cite{ulku2016multi, bekmezci2015location,ulku2019sharing} and it was not easy to test it in the GNU Radio environment. The system was tested in its final form, focusing on the physical layer; the difficulties encountered during development and further work were mentioned.

\subsection{Future Work}
Communication systems, by their nature, are systems that need to work efficiently and with performance. Especially in UAV systems, scenarios where mission success depends on communication efficiency arise due to the intense data exchange between UAVs or between the ground station and the UAV, the complexity of the mission, and the increase in the need for efficiency. In the literature, many system suggestions are made for this purpose. These studies address various sub-components for a more efficient system and develop suitable test environments. Communication systems used in real scenarios consist of very complex and expensive parts. However, since academic studies focus on narrower issues, it is too costly to develop the whole system. The goal may be to develop a routing algorithm or a better coding method. There are separate simulation environments, such as GNU Radio for the physical layer and ns-3 for higher layers (protocols). The problem here is that simulation environments cover only some of the communication systems, and they lack in creating realistic scenarios. Although there are API attempts to combine these two software platforms in the literature, they still need to be at the desired level. At the same time, it is not easy to develop a simulation system because software radio development is complex and requires a multidisciplinary perspective and experience. Even if it is possible to implement a physical layer purely software, it requires many tradeoffs. This tradeoff usually takes away from performance and system robustness. Some systems must respond and work at certain time intervals due to their functions; sometimes, it is impossible to provide this in software. For example, in the gr-ieee802-11 GNU Radio module used in this study, ACK, CTS, and RTS packets were timing out. To avoid this, running this part of the system on an FPGA may be preferable, but it will increase the cost. Another problem arises from implementing continuous system software in real-time. The increase in computational cost and the necessity of short response times are also difficulties in implementation.
In order to establish a more realistic system, software-defined radio hardware such as the USRP mentioned in this study can be used with the GNU Radio platform. The signal developed in software with this hardware can be spread to the environment in analog form through the transceivers on the SERPs. The most realistic environments and scenarios can be set up this way. Because of SDR's flexibility, different communication technologies can also be applied. The disadvantage is that each piece of hardware costs around 1,500 dollars, and if it is desired to establish a communication system between multiple devices, as in this study, the costs increase. Alternatively, protocols can be tested independently of the physical layer in simulation environments such as ns-3, but more may be needed for the physical and data link layers. As a result, it is important to set up realistic environments and scenarios, it may be necessary to implement a comprehensive communication system to do this, but it is not easy to implement. Developing an API for more holistic simulation environments or existing platforms will be beneficial in the long run. Although software radios provide flexibility during development, implementing a system requires expertise.

\bibliography{references}             

\appendix
\section{}

\begin{figure*}[htbp]
    \centering
    \includegraphics[width=\textwidth]{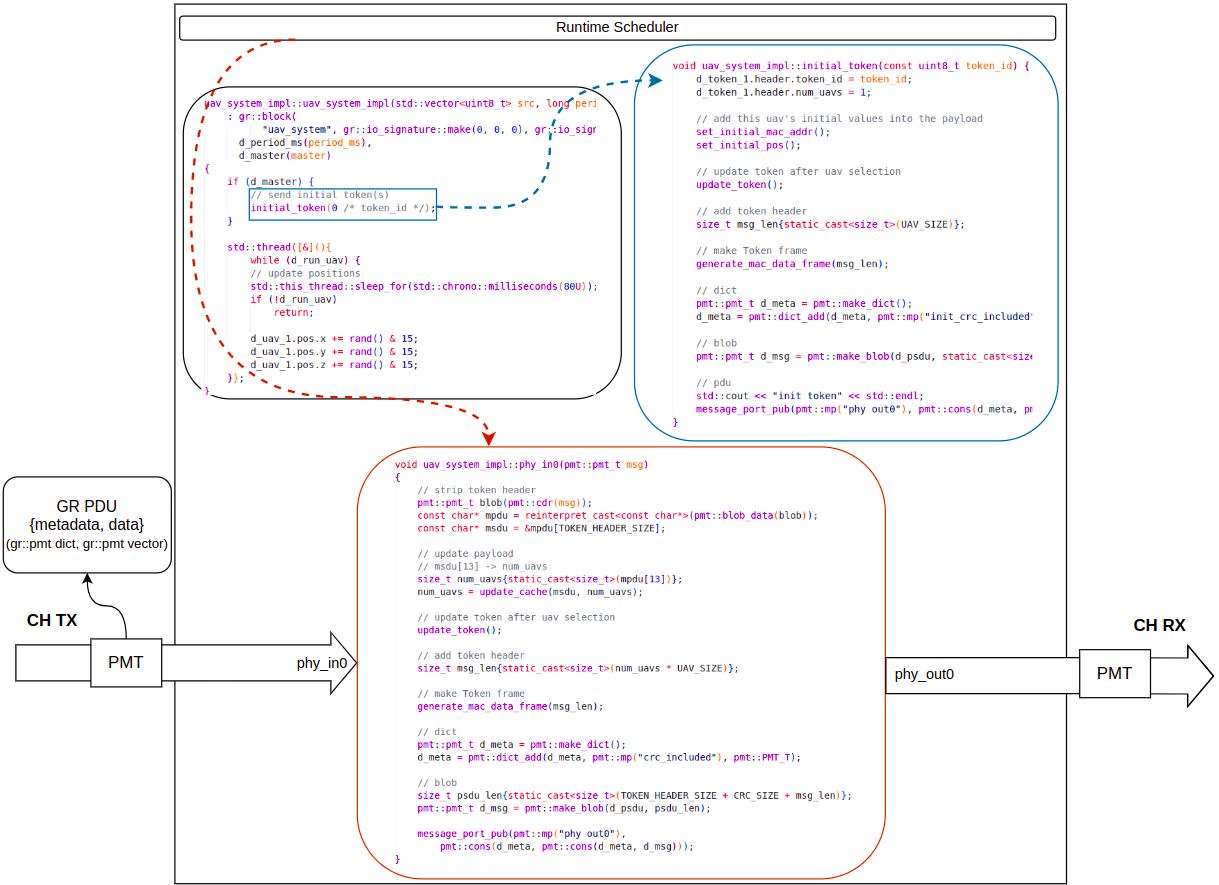}
    \captionsetup{justification=centering}
    \caption{Runtime overview of the system}
    \label{fig:system_over}
\end{figure*}

\end{document}